\documentclass{edp-conf}
\usepackage{graphicx}
\newcommand{\ltsimeq}{\raisebox{-0.6ex}{$\,\stackrel 
        {\raisebox{-.2ex}{$\textstyle <$}}{\sim}\,$}} 
\newcommand{\gtsimeq}{\raisebox{-0.6ex}{$\,\stackrel 
        {\raisebox{-.2ex}{$\textstyle >$}}{\sim}\,$}} 
\begin{document}

\TitreGlobal{AGN in their Cosmic Environment}

\title{Radio -- Optical Correlations in \\High-z Radio Galaxies and Quasars} 
\runningtitle{Radio -- Optical Correlations}

\author{Chris J. Willott} 
\address{Astrophysics, Department of Physics, Keble Road, Oxford, OX1
3RH, U.K.}
\email{cjw@astro.ox.ac.uk} 
%
\maketitle
\begin{abstract} 

The narrow emission line luminosities of radio-loud AGN are
well-correlated with their low-frequency radio luminosities.  This
correlation is linear and extends over at least four orders of
magnitude. The correlation is discussed in terms of a linear
relationship between the power in the radio-emitting jets and the
photoionizing (i.e. accretion) luminosity. Support for this theory
comes from a direct correlation between the optical continuum and
radio luminosities of steep-spectrum quasars. We discuss the
dependence of the observed luminosities of AGN on their black hole
masses and how close to the Eddington limit they are accreting.

\end{abstract}
%
\section{Introduction}

Despite all the observational effort which has gone into researching
extragalactic radio sources, there are still major uncertainties
regarding their fueling, radiation emission mechanisms and the
relationship between the active nuclei, host galaxies and larger-scale
environments. Since resolving the various central emission components
in most wavebands is still difficult, much progress has been made by
looking for correlations between observables. The multi-waveband
spectra of AGN give considerable scope for such studies. In this paper
I consider correlations between the low-frequency (i.e.
lobe-dominated) radio emission and optical quantities such as the
optical continuum and emission lines. These are particularly revealing
since, unlike many observables which are correlated due to
reprocessing of radiation from one form to another, the emission
regions and mechanisms are physically distinct and therefore telling
us something more fundamental about the way that active galaxies
are powered.

Radio sources can be crudely split into two groups with different
radio structures and luminosities: the low luminosity (monochromatic
151 MHz luminosity $L_{151} \ltsimeq 10^{25}$ WHz$^{-1}$sr$^{-1}$),
edge-darkened FRIs and the high luminosity, edge-brightened FRIIs
(also known as `classical doubles'). In this paper I will concentrate
on the FRII class of which much larger and more complete samples
exist.  Unified schemes have been extremely successful in describing
the different properties of powerful FRII radio galaxies and quasars
as due to a difference in the angle between the line-of-sight and the
jet axis (see Antonucci 1993 for a review of this subject). Both the
extended radio emission and (at least the majority of) the narrow line
emission of both quasars and radio galaxies are expected to be emitted
isotropically in such models. The cosmology assumed (except where
stated otherwise) is that $H_{\circ}=50~ {\rm km~s^{-1}Mpc^{-1}}$,
$\Omega_{\mathrm M} =1$, and there is zero cosmological constant.

\section{Optical continuum and radio emission in steep-spectrum quasars}

There have been varied results over the past 20 years with respect to
the existence of a correlation between the optical and radio
luminosities of radio-loud quasars due to:

\begin{itemize}

\item{Inhomogeneous samples with unknown
selection effects}

\item{Small complete samples with small range in
radio luminosity}

\item{Flat-spectrum quasars may have a correlation due to beaming of
optical and radio flux}

\end{itemize}

\begin{figure}
\includegraphics[width=12cm]{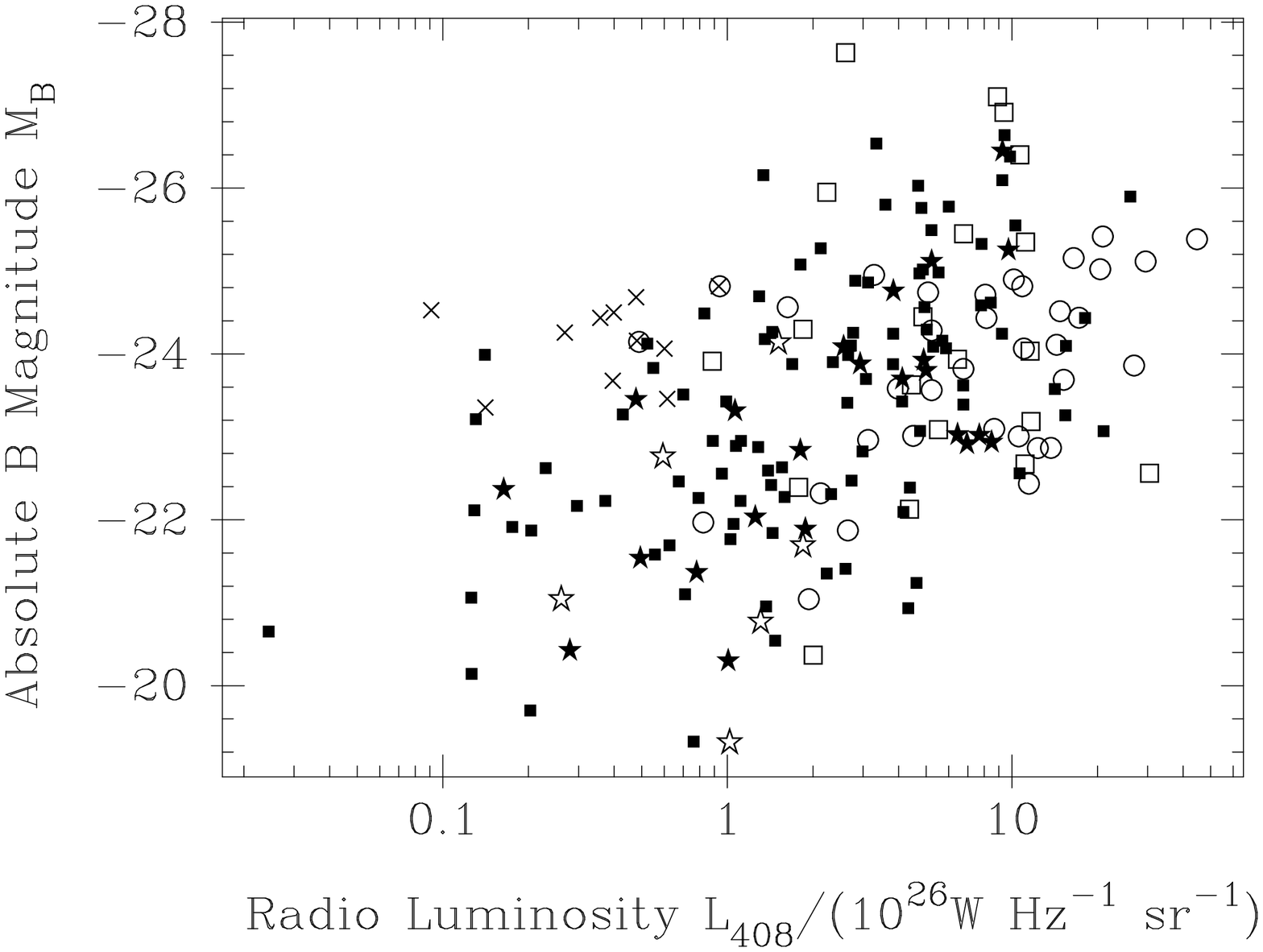}
\caption{Radio luminosity against absolute magnitude for
steep-spectrum quasars (SSQs) from the MAQS (squares), the MQS (stars)
(which both have flux-limits of S$_{408}$ = 0.95 Jy) and 3CRR
(circles). This plot is reproduced from Serjeant et al. (1998) and
assumes $H_{\circ}=100~ {\rm km~s^{-1}Mpc^{-1}}$.}
\end{figure}

\begin{figure}
\includegraphics[width=12cm]{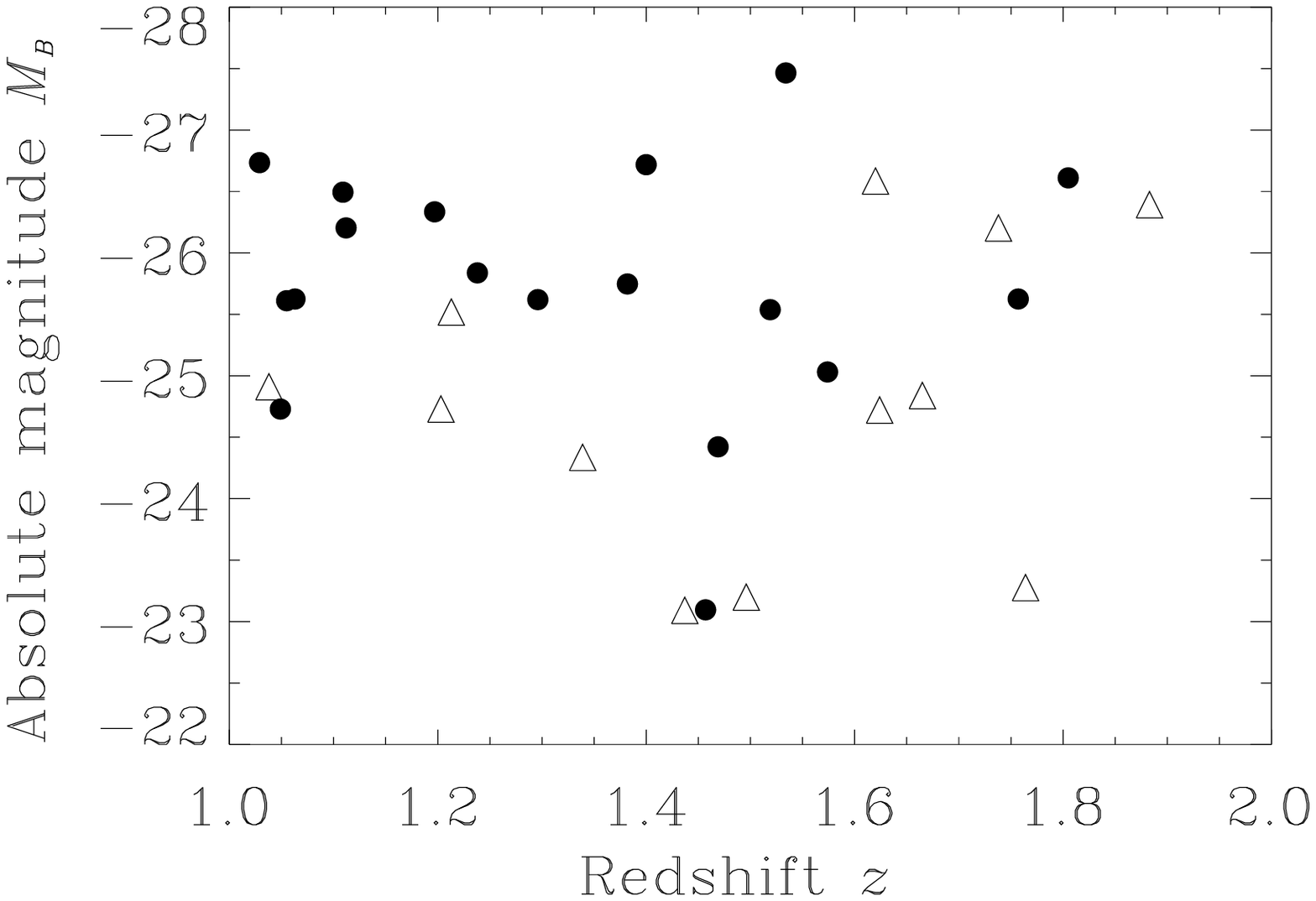}
\caption{Absolute magnitude against redshift for quasars in the 3CRR
(circles) and 7CRS (triangles) complete samples with $1<z<2$. Note the
vertical separation of the absolute magnitudes for the two samples.}
\end{figure}

One must also be able to differentiate between luminosity dependence
and evolution.  In any single flux-limited sample, the steepness of
the luminosity function ensures a tight luminosity-redshift
correlation so that the most luminous sources in the sample will
generally be those at the highest redshifts. Breaking this degeneracy
requires combining several samples selected in a similar manner with
different flux limits.

Browne \& Murphy (1987) found a correlation between the optical and radio
luminosities in a large, but inhomogeneous sample. Baker (1997) showed
that the optical continua of quasars with stronger radio cores are
brighter than those with weak cores and that emission line equivalent
widths anti-correlate with $R$ (the ratio of core to extended
flux). These observations can be explained as a combination of optical
beaming and anisotropic optical continuum - e.g. a disk or reddening.

To understand the situation properly requires samples of quasars
selected in an orientation-independent manner. Serjeant et al. (1998)
used low--frequency selected complete samples (178 \& 408 MHz) to show
that there is a correlation between the radio luminosity at
low--frequency and the optical continuum luminosity (Fig. 1).  The
Serjeant et al. best-fit relation gives L$_{\rm opt} \propto$ L$_{\rm
rad}^{0.6 \pm 0.1}$ with dispersion of 1.6 mag.

This correlation is also seen in a combination of 3CRR (178 MHz --
Laing, Riley \& Longair 1983) and 7CRS (151 MHz) complete (no optical
magnitude limit) SSQ samples (Willott et al. 1998). In the redshift
range 1 $<$ z $<$ 2, where there is the largest overlap between
quasars in the two samples, it is clear that the 3CRR sources (with
radio luminosities $\sim$ 25 times higher than 7CRS sources) are more
luminous in the optical (Fig. 2). The mean $M_B$ for 3CRR SSQs is
-25.9$\pm$ 0.2 mag, compared to -24.6 $\pm$ 0.3 for the 7CRS
SSQs. This shows a direct link between the optical continuum and
extended radio emission mechanisms for quasars.

A consequence is that radio fainter samples will contain
optically--fainter quasars. Therefore, despite the fact that all 3CRR
quasars are identified on the POSS-I plates (R $\ltsimeq 20$), this
should not be assumed for fainter radio samples (e.g.  only 60\% of
quasars in the 7CRS are detected on the POSS-I plates).

\section{Narrow-line emission in radio-loud AGN}

The narrow-line region (NLR) in radio sources is extended over several
kpc, beyond the proposed dusty torus which obscures the central
continuum source and broad-line region along certain
lines-of-sight. Therefore the narrow-line emission is believed to be
independent of the jet axis orientation.  Similar distributions of
narrow-line luminosities in samples of radio galaxies and quasars,
matched in extended radio luminosity, have been suggested as a
fundamental test of the unified schemes (e.g.\ Barthel 1989). In
general the emission line luminosities of well-matched samples of
radio galaxies and quasars are similar, in agreement with unified
schemes. There is some evidence for higher [OIII] $\lambda 5007$
luminosities in quasars than radio galaxies, but see Simpson (1998)
for a discussion of this and a possible resolution within the
framework of unified schemes.

\begin{figure}
\includegraphics[width=12cm]{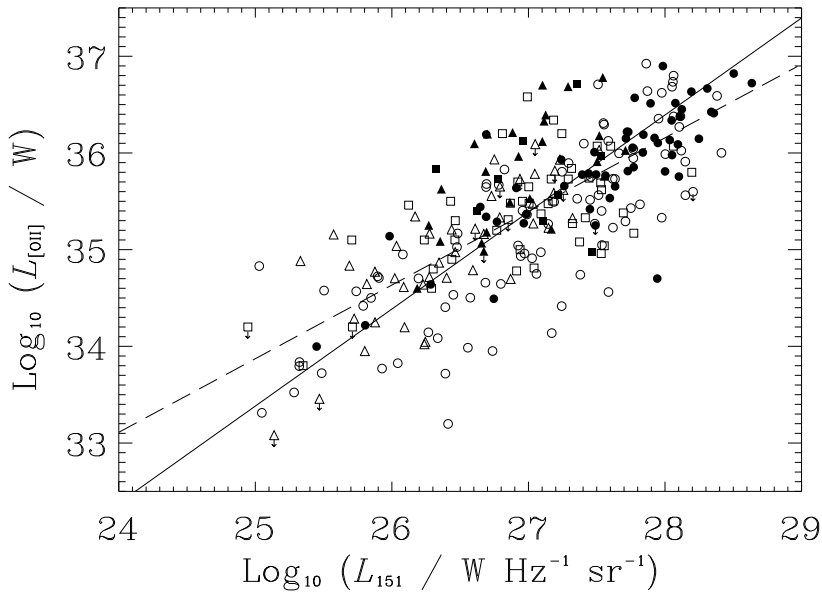}
\caption{The correlation between narrow [OII] emission line luminosity
and low-frequency radio luminosity for FRII radio sources in the 3CRR
(circles), 6CE (squares) and 7CRS (triangles) complete
samples. Quasars (and broad-lined radio galaxies) are shown as open
symbols and radio galaxies as filled symbols. Note that the [OII] flux
was not available for every object, so for about a third of these
objects another line was measured and the [OII] flux inferred using
the line ratios in McCarthy (1993).  The dashed line is a standard
least-squares linear fit to the data and the solid line a fit
minimizing the deviations along both axes. }
\end{figure}

There is a strong positive correlation between the extended radio
luminosities and narrow emission line luminosities of 3C radio sources
(Baum \& Heckman 1989; Rawlings et al. 1989). However, due to the
correlation between radio luminosity and redshift in any single
flux-limited sample, this could equally be interpreted as a
correlation between redshift and emission line luminosity. Using the
low-frequency selected 7CRS in combination with the 3CRR sample,
Willott et al. (1999) confirmed that the correlation is predominantly
with radio luminosity (although a weak residual correlation with
redshift remains). Fig. 3 plots the [OII] $\lambda 3727$ line
luminosity against low-frequency radio luminosity for these samples
(also included is the 6CE complete sample which has flux-limits in
between the 3CRR and 7CRS samples - see Rawlings, Eales \& Lacy
2000). As expected, the 6CE sources follow the same correlation as the
3CRR and 7C sources.

\begin{figure}
\hspace{1.3cm}
\vspace{0.3cm}
\includegraphics[width=10cm]{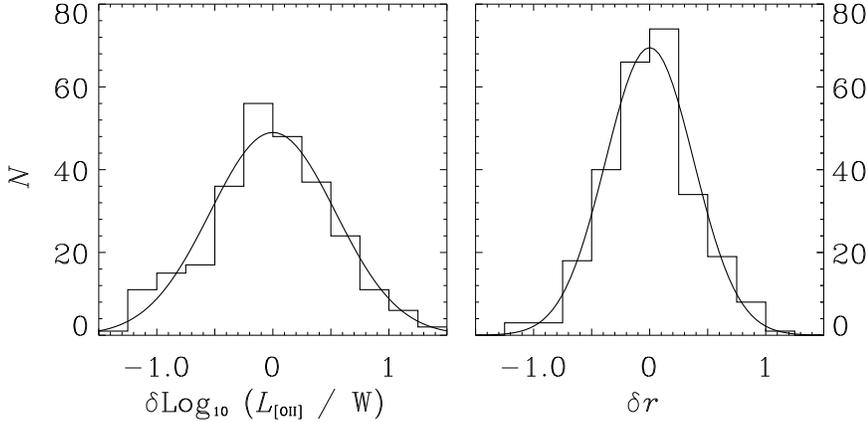}
\vspace{0.8cm}
\caption{Histograms of the scatter about the best-fit power-law
relationship between $L_{\rm [OII]}$ and $L_{151}$ (solid line in
Fig. 3). The left-hand side plots the scatter in $\log_{10} L_{\rm
[OII]}$ and the right-hand side shows the scatter in a direction
perpendicular to the best-fit line. Both distributions are close
approximations to gaussians.}
\end{figure}

\subsection{Slope}

Modeling the interpretation of this correlation requires the
relationship between the two variables to be quantified. There is no
apparent change in the slope of the correlation over the four orders
of magnitude shown in Fig. 3. To determine the slope a linear
least-squares fit is carried out in logarithm space. The best-fit
slope is $0.76 \pm 0.04$ (dashed line) -- not very different from that
of $0.79 \pm 0.04$ determined in Willott et al. (1999) without the 6CE
sample. The standard least-squares fit only minimizes the sum of the
square of the deviations of the model from the data along one axis --
in this case $\log_{10} L_{\rm [OII]}$. However, when one is dealing
with correlations between quantities such as two fluxes or two
luminosities which show scatter, the true relationship between the
quantities can only be obtained by minimizing the deviations along
{\rm both} axes simultaneously. As an example we find that by
performing a least-squares fit minimizing the square of the deviations
along the other axis (i.e. $\log_{10} L_{151}$), the slope of the
correlation is now $1.33 \pm 0.07$. The large difference between these
two values of the slope shows just how important this effect is in
this type of situation. A fit simultaneously minimizing the sum of the
squares of the deviations along both axes gives a slope of $1.00 \pm
0.04$ and is also shown on Fig. 3 (solid line). Thus we find a direct
proportionality between the narrow-line luminosity and radio
luminosity.

\subsection{Scatter}

This linear relationship between narrow line luminosity and radio
luminosity is intriguing, but we should never lose sight of the fact
that, as Fig. 3 shows, there is considerable scatter about this
relationship. Fig. 4 shows a histogram of the residuals between the
data and the best-fit slope of $1.00$. On the left-hand side are the
residuals in $\log_{10} L_{\rm [OII]}$. This is the scatter in the
values of $\log_{10} L_{\rm [OII]}$ for a given radio luminosity. The
distribution of residuals is close to gaussian (in log space) with
$\sigma=0.54$, i.e. $\pm 1 \sigma$ is one order of magnitude in
$L_{\rm [OII]}$. However, the true scatter about the best-fit
relationship is better defined in terms of the smallest deviation of
the data from the best-fit line (which we call $\delta r$). The values
of $\delta r$ are lower than the values of $\delta \log_{10} L_{\rm
[OII]}$ by a factor $\sin(\tan^{-1}(1.0/{\rm slope}))$. For the slope
of $1.00$ we find the residuals $\delta r$ are lower by a factor of
$0.7$ leading to a gaussian distribution with $\sigma=0.38$
(right-hand side of Fig. 4). There are various factors which can
account for the scatter in this correlation such as ranges in radio
source environments, NLR ionization states and NLR covering factors (see
Willott et al. 1999 for a more detailed discussion).

\section{Interpretation of radio -- optical correlations}

In the previous section we found that the correlation between narrow
emission line and radio luminosity extends over four orders of
magnitude with a linear slope. Interpretation of this correlation
requires 3 questions to be answered.

\subsection{What determines low-frequency radio luminosity?}

The radio luminosity of FRII radio sources at low-frequencies ($<1$
GHz) is dominated by the extended lobes . These lobes are fed by
relativistic particles accelerated in the hotspots which cool via
synchrotron radiation. However, the radio luminosity emitted is only a
small fraction of the total kinetic power of the jets, $Q_{\rm
jets}$. A much greater fraction is stored in the lobes and/or lost to
the environment via work done by the expanding radio source (Scheuer
1974). Models of the evolution of FRII sources lead to a nearly linear
relationship between $Q_{\rm jets}$ and $L_{151}$ (e.g. $Q_{\rm jets}
\propto L_{151}^{6/7}$ in the models of Rawlings \& Saunders 1991 and
Kaiser \& Alexander 1997).  These models also show that there is only
a weak negative dependence of the radio luminosity upon source size
(however, at large sizes the luminosity may undergo a rapid decrease
due to increased adiabatic and synchrotron losses -- Kaiser,
Dennett-Thorpe \& Alexander 1997; Blundell, Rawlings \& Willott 1999).

\subsection{What is the NLR ionization mechanism?}

The presence of high-excitation lines in radio galaxy and quasar
narrow-line spectra indicates that the active nucleus is the major
source of ionization in these objects. There are several possible
ionization mechanisms: photoionization, jet-cloud collisions and
shock-excitation. Photoionization by the central source is believed to
be the primary ionization mechanism for the majority of objects.
Studies of line ratios show shock-excitation is important for some
sources (Clark et al. 1997; Villar-Martin et al. 1999). However,
shocks have more effect on the morphology, kinematics and ionization
state than on the overall line luminosity. Interestingly, Best,
R\"ottgering \& Longair (2000) have found that the $z \sim 1$ 3CRR
radio galaxies with line emission matching the shock models also tend
to have the smallest radio sources. This implies that for some small
radio sources the passage of the jet through the narrow line region
can cause excess emission with a low-ionization state. Willott et
al. (1999) found a weak residual anti-correlation between emission
line luminosity and radio source linear size. This showed up as a lack
of sources with weak lines and small sizes, consistent with the
results of Best et al. However, in general photoionization is the most
likely mechanism for exciting the lines in most radio sources (see
discussion in Willott et al. 1999). In the ionization-bounded case,
where individual optically-thick clouds absorb and re-radiate all the
incident flux, the narrow-line luminosity is proportional to the
photoionizing quasar luminosity $Q_{\rm phot}$.

\subsection{What drives the luminosities?}

The low-frequency radio luminosity has a nearly linear relationship
with the bulk kinetic power in the jets $Q_{\rm jets}$. The narrow
emission line luminosity is proportional to the photoionizing quasar
luminosity $Q_{\rm phot}$. This allows us to interpret the observed
linear correlation between narrow line and radio luminosity as a
linear correlation between $Q_{\rm jets}$ and $Q_{\rm phot}$. Note
that the correlation between the optical and radio luminosities of
steep-spectrum quasars in Section 2 strongly supports this
interpretation. This rules out effects such as the environment
controlling both the radio and narrow-line luminosities. Although
there are large uncertainties in the absolute normalization between
radio luminosity and $Q_{\rm jets}$, Willott et al. (1999) found that
$0.05 \ltsimeq Q_{\rm jets} / Q_{\rm phot} \ltsimeq 1$, so that
the jet power is within about an order of magnitude of the UV
luminosity. Celotti et al. (1997) found a relationship between the
kinetic energy in parsec-scale jets and the broad emission line
luminosities of quasars. This study also showed that the ionizing
luminosity is of the same order of magnitude as the jet power.

What are the properties of the AGN that determine $Q_{\rm phot}$ and
$Q_{\rm jets}$? Although there is still debate about the exact UV
emission process in AGN -- accretion disk or free-free emission
(e.g. Siemiginowska et al. 1995), the energy source is the release of
energy as material falls into the deep potential well close to a
supermassive black hole. Using the fact that there is an upper limit
to the ionizing luminosity which can be emitted by a black hole of a
certain mass -- the Eddington limit $L_{\rm Edd}$ -- the luminosity
can be related to the black hole mass $M_{\rm BH}$ via $Q_{\rm phot}
\propto f_{\rm Edd} M_{\rm BH}$ where $f_{\rm Edd}$ is the fraction of
the Eddington luminosity being emitted, i.e.  $Q_{\rm phot}/L_{\rm
Edd}$ (we are neglecting here the infrared contribution to the
bolometric luminosity, which is of the same order as $Q_{\rm phot}$).
What determines the range of four orders of magnitude in $Q_{\rm
phot}$ in Fig. 3? Is it a similar size range in $M_{\rm BH}$ or
$f_{\rm Edd}$. The answer is that both play significant and probably
equal parts in determining the range of observed luminosities. The
objects with the highest values of $Q_{\rm phot}$ in Fig.  3 have
similar luminosities to the most luminous radio-quiet
(optically-selected) quasars. It seems likely that these objects are
emitting at approximately the Eddington limit ($f_{\rm Edd} \sim 1$)
leading to black hole masses $\sim 10^9 - 10^{10} M_{\odot}$.

\begin{figure}
\hspace{0.4cm}
\includegraphics[width=12cm]{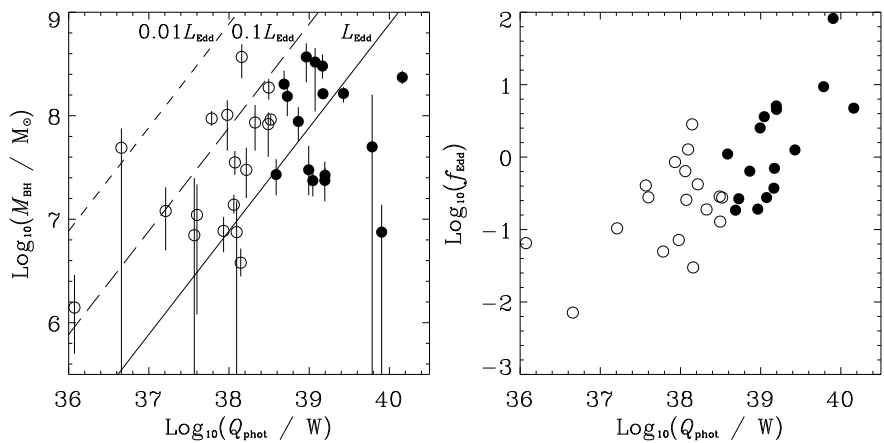}
\vspace{0.5cm}
\caption{Reverberation data of low-redshift, low-luminosity quasars
($M_B < -23$ -- filled symbols) and Seyfert 1 galaxies (open symbols)
from Kaspi et al. (2000) converted to our assumed value of the Hubble
constant $H_{\circ}=50~ {\rm km~s^{-1}Mpc^{-1}}$. The left-hand plot
shows black hole mass $M_{\rm BH}$ against $Q_{\rm phot}$ [calculated
assuming a bolometric correction of $Q_{\rm phot} = 10 \lambda
L_{\lambda}$(5100 \AA) as in Wandel, Peterson \& Malkan 1999]. The
right-hand plot shows $f_{\rm Edd}$ against $Q_{\rm phot}$ for the
same data (error bars omitted for clarity). Note the positive
correlation in this data. }
\end{figure}

Reverberation studies of low-redshift low-luminosity quasars give the
best estimates currently available of the range of black hole masses
in these objects. Kaspi et al. (2000) present data on the largest
sample of objects to date -- the results of which are shown in Fig. 5.
These objects (Seyfert 1s and quasars) have black hole masses in the
range $10^6-10^9 M_{\odot}$ and $f_{\rm Edd} \sim 0.01 - 1$.
There is evidence for a correlation between $Q_{\rm phot}$ and $f_{\rm
Edd}$, but no correlation between $M_{\rm BH}$ and $f_{\rm Edd}$. Thus
the range in AGN luminosity is due to similar size ranges in both
$M_{\rm BH}$ and $f_{\rm Edd}$. Note that the most luminous sources in
Fig. 3 are an order of magnitude more luminous than that yet sampled
by reverberation data and long-term ($\gtsimeq 10$ yr) monitoring of
high-luminosity quasars is needed to measure their broad-line region
sizes and black hole masses.

It is now well-established that many nearby galaxies (including our
own) contain large central mass concentrations, most probably in the
form of supermassive black holes (e.g. Richstone et al. 1998) and that
the masses of these black holes are linearly correlated with the
galaxy bulge masses (Magorrian et al. 1998). A similar correlation is
also seen for AGN with reverberation black hole masses (Wandel
1999). Roche, Eales \& Rawlings (1998) find that 6C radio galaxies at
$z\sim 1$ are smaller and fainter than 3CR galaxies (Best, Longair \&
R\"ottgering 1998) at the same redshift (which are a factor of about 6
times greater in radio luminosity). They conclude that there is indeed
a positive correlation between the radio luminosity and host galaxy
luminosity, which would seem to be caused by both quantities
correlating with black hole mass. The lack of such correlations at low
redshift (e.g. Eales et al. 1997) can be explained because the fuel
supply is more limited due to the virialization of galaxies and
clusters (e.g. Rees 1990) and the decrease in the galaxy merger rate
(Le F\`evre et al. 2000).

\section*{Acknowledgements}

Thanks to my collaborators Steve Rawlings, Katherine Blundell and Mark
Lacy for their work on the 7C Redshift Survey. Extra thanks to Steve
Rawlings for a critical reading of this manuscript.



\begin{thebibliography}{99}
\bibitem{142} Baker J.C., 1997, MNRAS, 286, 23  
\bibitem{152} Barthel P.D., 1989, ApJ, 336, 606
\bibitem{112} Baum S.A., Heckman T.M., 1989, ApJ, 336, 702 
\bibitem{160} Best P.N., Longair M.S., R\"ottgering H.J.A., 1998,
MNRAS, 295, 549
\bibitem{218} Best P.N., R\"ottgering H.J.A., Longair M.S., 2000,
MNRAS, 311, 23
\bibitem{217} Blundell K.M., Rawlings S., Willott C.J., 1999, AJ,
117, 677
\bibitem{158} Browne I.W.A., Murphy D., 1987, MNRAS, 226, 601 
\bibitem{341} Celotti A., Padovani P., Ghisellini G., 1997, MNRAS,
286, 415
\bibitem{253} Clark N.E. \etal\, 1997, MNRAS, 286,
558
\bibitem{175} Eales S.A. \etal\, 1997, MNRAS, 291, 593
\bibitem{137} Kaiser C.R., Alexander P., 1997, MNRAS, 286, 215
\bibitem{138} Kaiser C.R., Dennett-Thorpe J., Alexander P., 1997,
MNRAS, 292, 723
\bibitem{100} Kaspi S. \etal\, 2000, ApJ, 533, 631
\bibitem{309} Laing R.A., Riley J.M., Longair M.S., 1983, MNRAS,
204, 151 
\bibitem{156} Le F\`evre O. \etal\, 2000, MNRAS, 311, 565
\bibitem{138} Magorrian J. \etal\, 1998, AJ, 115, 2285
\bibitem{1}   McCarthy P.J., 1993, ARAA, 31, 639 
\bibitem{9}   Rawlings S., Saunders R., 1991, Nature, 349, 138 
\bibitem{19}  Rawlings S., Saunders R., Eales S.A., Mackay C.D., 1989,
MNRAS, 240, 701
\bibitem{355} Rawlings S., Eales S.A., Lacy M., 2000, MNRAS, submitted
\bibitem{15}  Rees M.J., 1990, Science, 247, 817
\bibitem{140} Richstone D., \etal\, 1998, Nature, 395, A14
\bibitem{168} Roche N., Eales S.A., Rawlings S., 1998, MNRAS, 297, 405
\bibitem{206} Scheuer P.A.G., 1974, MNRAS, 166, 513
\bibitem{359} Serjeant S. \etal\, 1998, MNRAS, 294, 494
\bibitem{359} Siemiginowska A. \etal\, 1995, ApJ, 454, 77
\bibitem{360} Simpson C., 1998, MNRAS, 297L, 39
\bibitem{228} Villar-Martin M., Tadhunter C.N., Clark N. E., 1997,
A\&A, 323, 21
\bibitem{227} Wandel A., 1999, ApJ, 519, L39
\bibitem{229} Wandel A., Peterson B.M., Malkan M.A., 1999, ApJ, 526,
579
\bibitem{13}  Willott C.J., Rawlings S., Blundell K.M., Lacy M., 1998,
MNRAS, 300, 625
\bibitem{14}  Willott C.J., Rawlings S., Blundell K.M., Lacy M., 1999,
MNRAS, 309, 1017


\end{thebibliography}
\end{document}